\documentclass[eqsecnum,aps,prb,twocolumn]{revtex4}
\usepackage{graphicx}
\usepackage{amsmath}
\usepackage{amsfonts}
\usepackage{amssymb}

\def\Dijm{\overline{D}_{ij}}
\def\dij{d_{ij}}

\newcommand{\Derm}[2]{\frac{{\rm d} \overline{#1}}{{\rm d} {#2}}}
\newcommand{\Dermp}[2]{\frac{{\rm d} \overline{#1}^+}{{\rm d} {#2^+}}}

\newcommand{\carDer}[2]{\left(\frac{{\rm d} \overline{#1}}{{\rm d} {#2}}\right)^{2}}
\newcommand{\carDerp}[2]{\left(\frac{{\rm d} \overline{#1}^+}{{\rm d} {#2^+}}\right)^{2}}

\newcommand{\proderr}[4]{\overline{\frac{\partial{#1}}{\partial {#2}}\frac{\partial{#3}}{\partial{#4}}}}

\newcommand{\proderrm}[4]{\frac{\partial \overline{#1}}{\partial{#2}}\frac{\partial \overline{#3}}{\partial{#4}}}

\newcommand{\derrydd}[3] {\frac{\partial^{2}{#1}}{\partial{#2} \,\partial{#3}}}

\begin{document}
\title{Reynolds number effect on the dissipation function in wall-bounded flows}
\author{F. Laadhari}
\email{faouzi.laadhari@univ-lyon1.fr}
\affiliation{Laboratoire de M\'ecanique des Fluides et d'Acoustique\\ 
Universit\'e de Lyon; Universit\'e Lyon 1; Ecole Centrale de Lyon; INSA de Lyon;\\CNRS, UMR  5509.\\36 Avenue Guy de Collongue, F-69134 Ecully, France}
\date{}
\begin{abstract}
The evolution with Reynolds number of the dissipation function, normalized by wall variables, is investigated using direct numerical simulation databases for incompressible turbulent Poiseuille flow in a plane channel, at friction Reynolds numbers up to $Re_\tau=2000$. DNS results show that the mean part, directly dissipated by the mean flow, reaches a constant value while the turbulent part, converted into turbulent kinetic energy before being dissipated, follows a logarithmic law. This result shows that the logarithmic law of friction can be obtained without any assumption on the mean velocity distribution. The proposed law is in good agreement with experimental results in plane-channel and boundary layer flows.
\end{abstract}
\maketitle
Despite extensive study, there remain significant questions about Reynolds number effects on wall-bounded flows. Earlier surveys of data,\cite{Gad-el-HaketBandyopadhyay94,FernholzetFinley96} indicated that Reynolds number effects are present in the near-wall region over a wide range of Reynolds number. The major difficulty in drawing firm conclusions is the accuracy of the measurements, which invariably are affected by spatial resolution and other near-wall measurement issues.

Direct numerical simulations (DNS) of turbulent flows provide detailed turbulence data that are free from such experimental ambiguities. During the past two decades, the investigations of wall bounded turbulent flows by DNS have provided considerable insights into both the statistical and structural characteristics of wall bounded turbulence. One of the most well-studied turbulent flows is the flow in a plane channel, which was simulated by Kim \textit{et al.}\cite{Kimal87} and by many others since.\cite{Moseral99,Laadhari02,Iwamotoal02}  Moderately high Reynolds number simulations have been recently performed and the results made available.\cite{Abeal04,HoyasetJimenez06}

The aim of the present study is to investigate the Reynolds number dependence of the energy dissipation function in a turbulent plane channel flow using the results of both available DNS databases and our own simulations conducted here in order to obtain a wider and more complete range of Reynolds number.
The simulation parameters of DNS cases considered are given in
Table \ref{tablesym}.
\begin{table}
\caption{Parameters of the turbulent plane channel DNS datasets used. Reynolds numbers: $R_{e_b}=U_b h/\nu$, $R_{e_\tau}=u_\tau h/\nu$.}
\label{tablesym}
\begin{ruledtabular}
\begin{tabular}{lcrrc}
&$R_{e_b}$&$R_{e_\tau}$&Symbols\\ \tableline
Present study
	&	1015	&	72	&\\
	&	1300	&	90	&\\
	&	1800	&	120	&\\
	&	2480	&	160	&\\
	&	2830	&	180	&{\Large$\circ$}\\
	&	3830	&	235	&\\
	&	11000	&	590	&\\
	&	20100	&	1000	&\\
	&	30600	&	1450	&\\
\hline
Moser \textit{et al.} (Ref. \onlinecite{Moseral99})
	&	2800	&	178	&\\
	&	6880	&	392	&$\bigtriangleup$\\
	&	10950	&	587	&\\
\hline
Hoyas and Jimenez (Ref. \protect\onlinecite{HoyasetJimenez06})
	&	10060	&	547	&\\
	&	18520	&	934	&$\bigtriangledown$\\
	&	43600	&	2003	&\\
\hline
Iwamoto \textit{et al.} (Ref. \onlinecite{Iwamotoal02})
	&	1610	&	109	&\\
	&	2290	&	150	&\\
	&	5020	&	298	&$\lozenge$\\
	&	6960	&	396 	&\\
	&	12140	&	643	&\\
\hline
Tanahashi \textit{et al.} (Ref. \onlinecite{Tanahashietal04})
	&	7030	&	400	&$\square$\\
	&	17390	&	792	&\\
\end{tabular}
\end{ruledtabular}
\end{table}

The present numerical simulations are based on a pseudo-spectral code using the Chebychev-tau formulation in the wall-normal direction ($x_2$) and Fourier expansion in the streamwise ($x_1$) and spanwise ($x_3$) directions where periodic boundary conditions are applied.\cite{GodeferdetLollini99}
 The number of Fourier/Chebychev modes was selected so that the energy spectra are at sufficiently small values at large wave numbers, particularly near the
wall.
The flow was driven by a constant streamwise pressure-gradient  $\partial \overline{P}/\partial x_1$.

The mean energy dissipation rate per unit volume $\phi$, for incompressible flow, is given by:\cite{Corrsin53}
$$
\phi = 2\mu\Dijm\Dijm+2\mu\overline{\dij\dij}
$$
using standard Cartesian tensor notation and summation on repeated
indices. $\mu$ is the dynamic viscosity, $\Dijm$ and $\dij$ are respectively the mean and fluctuating part of the velocity deformation tensor. The first term on the right-hand side represents the part directly dissipated by the mean flow $\phi_M$, and the second, $\phi_T$, the turbulent part. The two terms are usually decomposed as the sum of homogeneous and inhomogeneous parts as follows:

$$
\phi_M=\mu\proderrm{U_{i}}{x_{j}}{U_{i}}{x_{j}}+\mu\derrydd{\overline{U}_{i}\overline{U}_{j}}{x_{i}}{x_{j}},
\label{eq1}
$$
and
$$
\phi_T=\mu\proderr{u_{i}}{x_{j}}{u_{i}}{x_{j}}+\mu\derrydd{\overline{u_{i}u_{j}}}{x_{i}}{x_{j}}.
\label{eq2}
$$
For fully turbulent flow in a plane channel, the variation of mean values in the streamwise and spanwise directions  are zero. The mean velocity
reduces to the streamwise component $\overline{U}_1$ which, like the Reynolds stresses, depends only on the wall-normal position $x_2$.
Since the gradient of the Reynolds stresses is zero at the channel walls, located at $x_2=\pm h$, the dissipation function $\Phi$, defined as the integral over the channel cross-section of the mean energy dissipation rate, is given by
\begin{equation*}
\Phi=\int_{-h}^{+h}\left[\mu\carDer{U_{1}}{x_{2}}+\mu\proderr{u_{i}}{x_{j}}{u_{i}}{x_{j}}\right]{\rm d}x_2
\end{equation*}

For the same reason, the turbulent kinetic energy equation (Ref.~\onlinecite{Karman37}, Eq. 5) integrated across the channel, shows that the turbulent contribution to the dissipation function is equal to the integral of the turbulent kinetic energy production:
$$
\Phi_T=\int_{-h}^{+h}\mu\proderr{u_{i}}{x_{j}}{u_{i}}{x_{j}}{\rm d}x_2=\int_{-h}^{+h}-\rho\overline{u_1u_2}\Derm{U_{1}}{x_{2}}{\rm d}x_2,
$$
where $\rho$ is the fluid density and $\overline{u_1u_2}$ the Reynolds shear stress, and finally
$$
\Phi=\int_{-h}^{+h}\left[\mu\carDer{U_{1}}{x_{2}}-\rho\overline{u_1u_2}\Derm{U_{1}}{x_{2}}\right]{\rm d}x_2.
$$
Using the integrated streamwise mean momentum equation:
\begin{equation*}
\mu\Derm{U_{1}}{x_{2}}-\rho\overline{u_1u_2}=-\tau_w\frac{x_2}{h},
\end{equation*}
where $\tau_w$ is the mean wall-shear stress, the dissipation function can now be easily evaluated:
$$
\Phi=\int_{-h}^{+h}-\tau_w\frac{x_2}{h}\Derm{U_{1}}{x_{2}}{\rm d}x_2.
$$
The integration by parts with the no-slip conditions at the walls leads to
$$
\Phi=\frac{\tau_w}{h}\int_{-h}^{+h}\overline{U_{1}}{\rm d}x_2=2\tau_wU_{b}
$$
where $U_{b}$ is the bulk velocity. This is the classical relation for the loss of power in a duct, which is equal to the product of the mean pressure gradient  $-\partial \overline{P}/\partial x_1=\tau_w/h$ by the flow rate $2hU_{b}$.

Note that an identical relation can be obtained for the turbulent flow in a circular pipe (see for example Ref. \onlinecite{Laufer53}), while for the turbulent boundary layer on a flat plate, the dissipation function is related to the streamwise variation of the mean and turbulent kinetic energy (see Eq. 3.15 in Ref. \onlinecite{Rotta53}).

In a dimensionless form, the dissipation function is now given by
\begin{equation*}
 \Phi^+=\int_{-h^+}^{+h^+}\left[\carDerp{U_{1}}{x_{2}}-\overline{u_1u_2}^+\Dermp{U_{1}}{x_{2}}\right]{\rm d}x^+_2=2U_{b}^+,
\end{equation*}
where the superscript $(+)$ denotes normalization by the friction velocity $u_\tau=\sqrt{\tau_w/\rho}$ and the kinematic viscosity $\nu$.

\begin{figure}
\includegraphics[angle=90,scale=0.32]{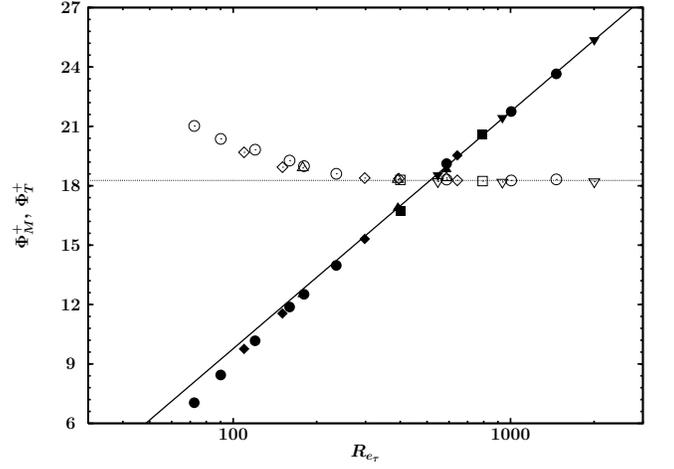}
\caption{Evolution with the friction Reynolds number $R_{e_\tau}$ of the mean shear $\Phi_M^+$ (open symbols) and turbulent $\Phi_T^+$ (filled symbols) contributions to the nondimensionalized dissipation function. --------, [Eq. (\ref{phit})]. Symbols in Table
\ref{tablesym}.}
\label{intSdeuxhplus}
\end{figure}

Figure \ref{intSdeuxhplus} shows, in a semilogarithmic plot, the evolution with $R_{e_\tau}$ of $\Phi^+_M$ and $\Phi^+_T$, the mean and turbulent contributions, respectively. This figure highlights that for  $R_{e_\tau}>500$, when $\Phi^+_T$ becomes greater than $\Phi^+_M$, $\Phi^+_M$ reaches a constant value and $\Phi^+_T$  follows a well defined logarithmic evolution. The DNS results give the two relations
\begin{eqnarray}
\Phi^+_M&=&18.27\nonumber\\
\Phi^+_T&=&5.2\,\ln \frac{R_{e_\tau}}{512} +18.27
\label{phit}
\end{eqnarray}

\begin{figure}
\includegraphics[angle=90,scale=0.32]{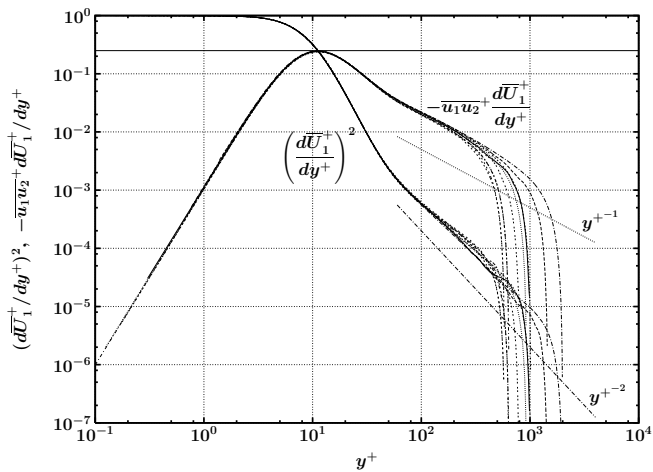}
\caption{Profiles of the mean-square velocity gradient and the mean production for $R_{e_\tau}>500$.}
\label{prod}
\end{figure}

\begin{figure}
\includegraphics[angle=90,scale=0.32]{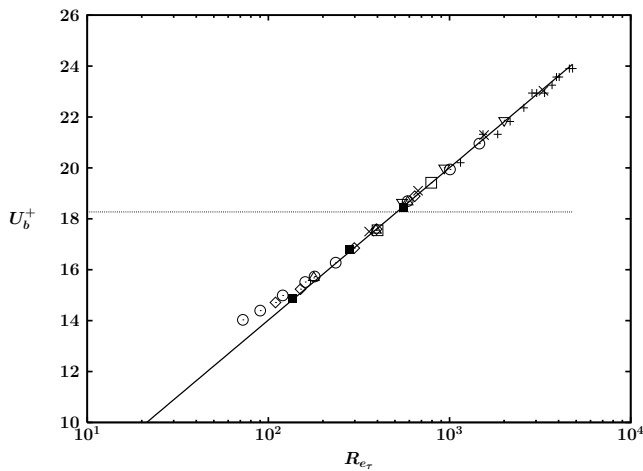}
\caption{Reynolds number evolution of $U_b^+$, the bulk velocity normalized by friction velocity. Experimental data: ($+$), Zanoun {\it et al.} (Ref. \onlinecite{Zanounetal03}); ($\times$), Bakken {\it et al.} (Ref. \onlinecite{Bakkenetal05}); ({\small$\blacksquare$}) DNS results of Spalart (Ref. \onlinecite{Spalart88}); --------, [Eq. (\ref{Ubplus})]. Other symbols are in Table
\ref{tablesym}.}
\label{Umplus}
\end{figure}

\begin{figure}
\includegraphics[angle=90,scale=0.32]{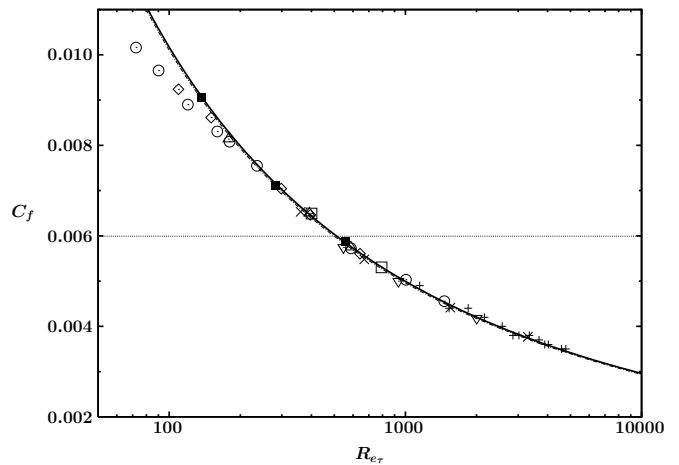}
\caption{Skin-friction coefficient. --------, logarithmic
friction law from Eq. (\ref{Ubplus}); {\tiny$-\cdot-\cdot-\cdot-$} best fit of boundary layer measurements of {\"{O}}sterlund {\it et al.} (Ref. \onlinecite{Osterlund00}). Other symbols are the same as in Fig. \ref{Umplus}.}
\label{Cfm}
\end{figure} 

 These results are not too surprising according to the profiles of the squared mean velocity gradient and the mean production for \mbox{$R_{e_\tau}>500$} shown in figure \ref{prod}. In the region \mbox{$y^+=h^+-|x_2^+|\leqslant 30$} the profiles exhibit universal behavior. Above this position, the square of the mean velocity gradient decreases faster than \mbox{$y^{+^{-1}}$}, giving a negligible contribution to the integral, while the mean production decay is close to \mbox{$y^{+^{-1}}$} in a region whose extent increases with Reynolds number.

The skin friction coefficient $C_f$, based on the bulk velocity is given by:
\begin{equation}
U_b^+=\sqrt{\frac{2}{C_f}}=2.6\ln R_{e_\tau} +2.05
\label{Ubplus}
\end{equation}
and might be extrapolated to arbitrarily large Reynolds numbers.
This logarithmic law is compared in Fig. \ref{Umplus} to the channel-flow experimental data of Zanoun {\it et al.},\cite{Zanounetal03} and  Bakken {\it et al.}\cite{Bakkenetal05}. Agreement with the data is within $\pm1\%$ for \mbox{$R_{e_\tau}\geqslant500$}.

This finding shows that for \mbox{$R_{e_\tau}>500$}, the dissipation function is free from low Reynolds number effects and corroborates the classical logarithmic law of friction without the assumption on the mean velocity profile used in the classical Prandtl-von~K\'arm\'an analysis,\cite{Karman34,Millikan38} based on the logarithmic law of the wall
$$
\overline{U}_1^+=\frac{1}{\kappa}\ln y^+ +A
$$
and the defect law
$$
\overline{U}_c^+-\overline{U}_1^+=-\frac{1}{\kappa}\ln \frac{y}{h} +B,
$$
where $\overline{U}_c$ is the mean centerline velocity.
The logarithmic skin friction law is obtained by assuming very large Reynolds number and combining the two equations
(see Ref. \onlinecite{Schlichting}, page 573):
$$
U_b^+=\frac{1}{\kappa}\ln R_{e_\tau} +A-\frac{1}{\kappa}.
$$
Hence, the factor $2.6$ in relation (\ref{Ubplus}) corresponds to a von~K\'arm\'an constant $\kappa=0.385$ which is close to the experimental value of $0.38$ obtained by {\"{O}}sterlund {\it et al.}\cite{Osterlund00} in a zero pressure-gradient turbulent boundary layer. However, a small discrepancy with experiment is found for the value of the additive constant $A$ since relation (\ref{Ubplus}) leads to $A=4.65$ while {\"{O}}sterlund {\it et al.}\cite{Osterlund00} obtained $A=4.1$.

Another noteworthy feature is the good agreement of the skin-friction law obtained from Eq. (\ref{Ubplus}) with the boundary layer skin-friction coefficient based on
 \mbox{$U_b=\left(1-\delta_1/\delta \right)U_\infty$}, where $U_\infty$, $\delta$ and $\delta_1$ are respectively the free-stream velocity, the boundary-layer thickness and the displacement thickness. This is obvious from Fig. \ref{Cfm} where the experimental data of {\"{O}}sterlund {\it et al.}\cite{Osterlund00} and the DNS results of Spalart\cite{Spalart88} are compared to the present relation. This finding requires further analysis since the skin-friction coefficient in this flow is given by the streamwise mean-momentum variation.

To summarize, the logarithmic law of friction for plane channel flow is the direct consequence of the logarithmic evolution of the dissipation function and more precisely of the turbulent part. The proposed law of friction is found to be in good accordance with boundary layer results.
A more detailed study including numerical and accurate measurements would be useful, particularly for Poiseuille pipe flow.
\begin{acknowledgments}
The computations were performed on the IBM SP4 parallel supercomputers at the CINES (Centre Informatique National
de l' Enseignement Sup\'erieur). We acknowledge the support of CINES
for providing the computational resources.
\end{acknowledgments}

\bibliographystyle{osa}

\begin{thebibliography}{10}

\bibitem{Gad-el-HaketBandyopadhyay94}
M. {Gad-el-Hak} and P.~R. Bandyopadhyay, ``Reynolds number effects in
  wall-bounded turbulent flows,'' Appl. Mech. Rev. {\bf 47,} 307--365 (1994).

\bibitem{FernholzetFinley96}
H.~H. Fernholz and P.~J. Finley, ``The incompressible zero-pressure-gradient
  turbulent boundary layer: An assessment of the data,'' Prog. Aerospace Sci.
  {\bf 32,} 245--311 (1996).

\bibitem{Kimal87}
J. Kim, P. Moin, and R.~D. Moser, ``Turbulence statistics in fully developed
  channel flow at low Reynolds Number,'' J.{\,}Fluid Mech. {\bf 177,} 133
  (1987).

\bibitem{Moseral99}
R.~D. Moser, J. Kim, and N.~N. Mansour, ``Direct numerical simulation of
  turbulent channel flow up to $Re_\tau=590$,'' Phys. Fluids {\bf 11,} 943--945
  (1999).

\bibitem{Laadhari02}
F. Laadhari, ``On the evolution of maximum turbulent kinetic energy production
  in a channel flow,'' Phys.{\,}Fluids {\bf 14,} L65--L68 (2002).

\bibitem{Iwamotoal02}
K. Iwamoto, Y. Suzuki, and N. Kasagi, ``Reynolds number effect on wall
  turbulence: toward effective feedback control,'' Int. J.{\,}Heat and Fluid
  Flow {\bf 23,} 678 (2002).

\bibitem{Abeal04}
H. Abe, H. Kawamura, and Y. Matsuo, ``Surface heat-flux fluctuations in a
  turbulent channel flow up to $Re_\tau=1020$ with Pr=0.025 and 0.71,'' Int. J.
  Heat and Fluid Flow {\bf 25,} 404--419 (2004).

\bibitem{HoyasetJimenez06}
S. Hoyas and J. Jim\'enez, ``Scaling of the velocity fluctuations in turbulent
  channels up to $Re_\tau=2003$,'' Phys. Fluids {\bf 18,} 011702 1--4 (2006).

\bibitem{GodeferdetLollini99}
F.~S. Godeferd and L. Lollini, ``Direct numerical simulations of turbulence
  with confinement and rotation,'' J.{\,}Fluid Mech. {\bf 393,} 257--308
  (1999).

\bibitem{Corrsin53}
S. Corrsin, ``Interpretation of viscous terms in the turbulent energy
  equation,'' J.\,Aeronautical Sciences. {\bf 20,} 853--854 (1953).

\bibitem{Karman37}
T. {Von~K\'arm\'an}, ``The Fundamentals of the Statistical Theory of
  Turbulence,'' J.{\,}Aeronaut. Sci. {\bf 4,} 131--138 (1937).

\bibitem{Laufer53}
J. Laufer, ``The structure of turbulence in fully developed pipe flow,''
  Technical Note 2954, NACA (1953) .

\bibitem{Rotta53}
J. Rotta, ``On the theory of the turbulent boundary layer,'' TM 1344, NACA
  (1953) .

\bibitem{Zanounetal03}
E.-S. Zanoun, F. Durst, and H. Nagib, ``Evaluating the law of the wall in
  two-dimensionnel fully developped turbulent channel flows,'' Phys.{\,}Fluids
  {\bf 15,} 3079--3089 (2003).

\bibitem{Bakkenetal05}
O.~M. Bakken, P. Krogstad, A. Ashrafian, and H.~I. Andersson, ``Reynolds number
  effects in the outer layer of the turbulent flow in a channel with rough
  walls,'' Phys.{\,}Fluids {\bf 17,} 065101 (2005).

\bibitem{Karman34}
T. {Von~K\'arm\'an}, ``Turbulence and skin friction,'' J.{\,}Aeronaut. Sci.
  {\bf 1,} 1--20 (1934).

\bibitem{Millikan38}
C.~B. Millikan, ``A critical discussion of turbulent flows in channels and
  circular tubes,'' In {\em Fifth International Congress of Applied Mechanics},
    (1938).

\bibitem{Schlichting}
H. Schlichting, {\em Boundary-Layer Theory}, 7th ed. (McGraw-Hill, New York,
  1979).

\bibitem{Osterlund00}
J.~M. {\"{O}}sterlund, A.~V. Johansson, H.~M. Nagib, and M.~H. Hites, ``A note
  on the overlap region in turbulent boundary layers,'' Phys.{\,}Fluids {\bf
  12,} 1--4 (2000).

\bibitem{Spalart88}
P.~R. Spalart, ``Direct simulation of the turbulent boundary layer up to
  R=1410,'' J.{\,}Fluid Mech. {\bf 187,} 61--98 (1988).

\bibitem{Tanahashietal04}
M. Tanahashi, S.-J. Kangand, T. Miyamoto, S. Shiokawa, and T. Miyauchi,
  ``Scaling Law of Fine Scale Eddies in Turbulent Channel Flows up to
  $Re_\tau$=800,'' Int. J. Heat and Fluid Flow {\bf 25,} 331--340 (2004).

\end{thebibliography}

\end{document}